\documentclass{PoS}
\usepackage{url}
\title{The VOICE Survey : VST Optical Imaging of the CDFS and ES1 Fields}
\ShortTitle{The VOICE Survey}
\author{\speaker{M. Vaccari}$^{,1,2}$, G. Covone$^{3}$, M. Radovich$^{4}$, A. Grado$^{5}$, L. Limatola$^{5}$, M.T. Botticella$^{5}$, E. Cappellaro$^{4}$, M. Paolillo$^{3}$, G. Pignata$^{6}$, D. De Cicco$^{3}$, S. Falocco$^{3}$, L. Marchetti$^{1,2}$, M. Brescia$^{5}$, S. Cavuoti$^{5}$, G. Longo$^{3}$, M. Capaccioli$^{3}$, N. Napolitano$^{5}$, P. Schipani$^{5}$\\
           1) Department of Physics \& Astronomy, University of the Western Cape, Robert Sobukwe Road, 7535 Bellville, Cape Town, South Africa\\
           2) INAF - Istituto di Radioastronomia, via Gobetti 101, 40129 Bologna, Italy\\
           3) Department of Physics, University of Napoli "Federico II", via Cinthia 9, 80126 Napoli, Italy\\
           4) INAF - Osservatorio Astronomico di Padova, vicolo dell'Osservatorio 5, 35122 Padova, Italy\\
           5) INAF - Osservatorio Astronomico di Capodimonte, via Moiariello 16, 80131 Napoli, Italy\\
           6) Departamento de Ciencias Fisicas, Universidad Andres Bello, Avda. Republica 252, Santiago, Chile\\
           E-mail: \email{mattia@mattiavaccari.net}}
\abstract{We present the VST Optical Imaging of the CDFS and ES1 Fields (VOICE) Survey, a VST INAF Guaranteed Time program designed to provide optical coverage of two 4 deg$^2$ cosmic windows in the Southern hemisphere. VOICE provides the first, multi-band deep optical imaging of these sky regions, thus complementing and enhancing the rich legacy of longer-wavelength surveys with VISTA, Spitzer, Herschel and ATCA available in these areas and paving the way for upcoming observations with facilities such as the LSST, MeerKAT and the SKA. VOICE exploits VST's OmegaCAM optical imaging capabilities and completes the reduction of WFI data available within the ES1 fields as part of the ESO-Spitzer Imaging Extragalactic Survey (ESIS) program providing $ugri$ and $uBVR$ coverage of 4 and 4 deg$^2$ areas within the CDFS and ES1 field respectively. We present the survey's science rationale and observing strategy, the data reduction and multi-wavelength data fusion pipeline. Survey data products and their future updates will be released at \url{http://www.mattiavaccari.net/voice/} and on CDS/VizieR.}
\FullConference{4th Annual Conference on High Energy Astrophysics in Southern Africa\\
		25-27 August, 2016\\
		Cape Town, South Africa}
                \bigskip
                \hrule
                \bigskip
\begin{document}
\pagenumbering{arabic}
\setcounter{page}{2}
%
%
%
%
\section{Introduction}
The formation and evolution of galaxies remain among the most intriguing open questions in observational astronomy, with their two fundamental facets of the formation of stars and the assembly of stellar mass and how they depend on environment and nuclear activity (e.g. \cite{Kauffmann2003,Kauffmann2004}). In order to understand the physical mechanisms driving these processes, we need accurate estimates of stellar mass (SM) and star-formation rate (SFR) for large, representative galaxy samples over a significant fraction of cosmic time. On the other hand, understanding how galaxy properties are shaped by their local environment requires large volumes adequately probing the large scale structure. In fact, tracing both star formation histories and stellar mass build-up across cosmic times and environments sets powerful constraints on theoretical scenarios of galaxy formation (e.g. \cite{Bower2006,Bower2010}).

Deep optical imaging surveys have long been the workhorse of galaxy formation and evolution studies. The advent of digital detectors in particular has enabled wide (e.g., SDSS, \cite{York2000}) and deep (e.g., HDF, \cite{Williams1996}) surveys, shedding light on galaxy properties from the local to the high-redshift Universe. Optical imaging offered not only excellent image quality but also allowed to observe galaxy rest-frame UV emission and identify high-redshift galaxies through the drop-out technique \cite{Steidel1996,Verma2007}.

Over the past decade, advances in infrared detectors for both ground-based and space-based telescopes such as VISTA (\cite{VISTA2004}), Spitzer (\cite{Spitzer2004}) and Herschel (\cite{Herschel2010}) have shifted the emphasis toward longer wavelengths. Still, even in the era of multi-wavelength astronomy, optical data retain a fundamental role in identifying and characterizing high-redshift sources detected in the infrared. Deep optical surveys of the southern sky, in particular, have somewhat lagged behind infrared surveys due to the lack of new optical imagers since the introduction of ESO's WFI and NOAO's MOSAIC in the late nineties.

Over the last few years, however, OmegaCAM \cite{Kuijken2011} on the 2.6~meter VLT Survey Telescope (VST, \cite{Capaccioli2011}) on Cerro Paranal and the Dark Energy Camera (DECAM, \cite{Flaugher2012}) on the Blanco 4~meter telescope on Cerro Tololo have started operations, enabling for the first time deep and wide optical imaging surveys of the southern sky on a par with their northern counterparts and well-matched with the sensitivity of infrared surveys. Optical multi-band imaging allows us to identify and determine accurate photometric redshifts, stellar masses and star formation rates, shedding light on galaxy stellar mass assembly and thus on the cosmic star formation history. The times are therefore ripe for deep and wide surveys of the southern sky.

The VST Optical Imaging of the CDFS and ES1 Fields (VOICE) Survey is a VLT Survey Telescope (VST; \cite{Capaccioli2011}) program using the 1 deg$^2$ OmegaCAM imager to carry out systematic optical imaging of two key cosmic windows in the southern hemisphere. Namely, VOICE targets two 4 deg$^2$ areas centered on the Chandra Deep Field South (CDFS; RA = 03:32:30 , DEC= -27:48:30) and on the European Large Area ISO Survey South 1 (\cite{Oliver2000}, ELAIS-S1; RA=00:34:45; DEC=-43:28:00), respectively. 
Hereafter, when emphasis is required on the VOICE footprints vs coverage at other wavelengths we will refer to these two 4 deg$^2$ fields as VOICE-CDFS and VOICE-ES1.

These two sky regions have been surveyed by a number of facilities, including VISTA/VIDEO (PI: Jarvis, \cite{Jarvis2013}), Spitzer-Warm/SERVS (PI: Lacy, \cite{Mauduit2012}), Spitzer-Cold/SWIRE (PI: Lonsdale, \cite{Lonsdale2003}), Herschel/HerMES (PI: Oliver, \cite{Oliver2012}) and ATCA/ATLAS (PI: Norris, \cite{Norris2006}, \cite{Middelberg2008}, \cite{Franzen2015}) producing a wealth of data from the X-ray to the radio. As such, they are also prime targets for future southern surveys by facilities such as the LSST, MeerKAT and the SKA. In particular, the VOICE CDFS field will be targeted by the deepest radio observations to date with the Square Kilometer Array South African Precursor Telescope MeerKAT \cite{Jonas2009} as part of the LADUMA HI survey and the MIGHTEE Continuum Survey. However, a uniform optical coverage at the depth required to identify a large fraction of infrared and radio sources has long been missing, thus motivating the choice of these two fields for VST observations. Obtaining the best multi-wavelength data in this field was therefore key for South Africa to best exploit MeerKAT and SALT for galaxy evolution studies \cite{Jarvis2012}.


%
The science rationale of the survey span a wide range of topics. Multi-epoch observations of well-studied fields allow us to study the rate of SN explosions and the properties of the SN-host population cite{Cappellaro2015,Botticella2017} as well as the low-luminosity AGN population discovered by means of their optical variability \cite{DeCicco2015,Falocco2015}.
Deep high-fidelity optical images will also allow us to study galaxy groups and clusters at intermediate and high redshift. At intermediate redshift, the excellent data quality and depth will also allow us to study galaxy clusters as well as reconstruct dark matter maps via weak lensing (Fu et al., in prep). At high redshift ($z \sim 1.5$) galaxy cluster candidates have been identified using VST-VISTA-Spitzer data, and a VLT/KMOS spectroscopic follow-up program has been carried out (Mei et al. in prep). 

\section{Survey Design and Observations}\label{design.sec}
The main goal of the VOICE project is to provide deep and uniform optical imaging on two southern cosmic windows targeted by surveys such as VIDEO, SERVS, SWIRE, HerMES and ATLAS, where the available optical data were not adequate for their full scientific exploitation for galaxy formation and evolution studies. The detailed footprint of the sky areas imaged by VST was chosen to best match the regions observed by VIDEO, SERVS and ATLAS and the VST field of view.
%
%
%
%

An earlier effort to obtain deep optical data over the ES1 sky region was performed with the ESO-Spitzer wide-area Imaging Survey (ESIS, PI: Franceschini, \cite{Berta2006,Berta2008}, ESO Large Program ID 168.A-0322). The ES1 area was first identified as the absolute minimum of the Galactic emission at 100 $\mu$m in the Southern sky. It was first covered at 15 and 90 $\mu$m by the European Large-Area ISO Survey (ELAIS, \cite{Oliver2000,Rowan-Robinson2004}) and later in four IRAC and three MIPS bands by the Spitzer Wide-Area Infrared Extragalactic Survey (SWIRE, \cite{Lonsdale2003}). ESIS was thus designed with the specific goal to identify and characterize ELAIS and SWIRE sources as done by \cite{Gonzalez-Solares2011} in Northern ISO/Spitzer fields.

The original ESIS proposal included WFI $BVR$ and VIMOS $Iz$ imaging over about 5.5 deg$^2$. Eventually, the full 5.5 deg$^2$ were observed in $BVR$ with WFI, whereas only 4 deg$^2$ and 1 deg$^2$ were observed with VIMOS in $I$ and $z$, respectively. However, only a fraction of the collected WFI data have been reduced and analyzed so far. \cite{Berta2006} presented and released images and catalogs based on a subset of the WFI $BVR$ dataset covering 1.5 deg$^2$, while the VIMOS $Iz$ dataset was reduced in its entirety and presented by \cite{Berta2008}.

\begin{figure*}
\centering
\includegraphics[width=0.45\textwidth]{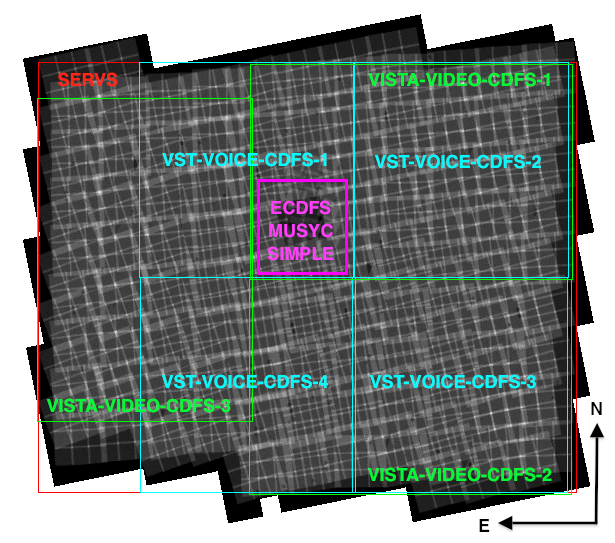}
\includegraphics[width=0.45\textwidth]{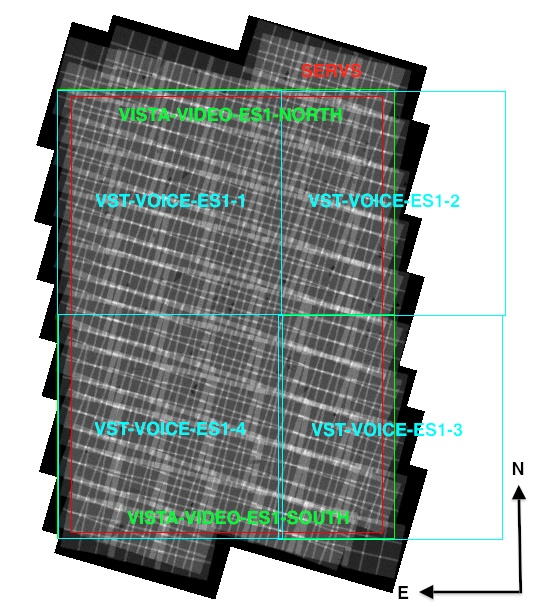}
\caption{VOICE Sky Coverage in the context of Multi-Wavelength Observations}
\label{fields.fig}
\end{figure*}

\begin{figure*}
\centering
\includegraphics[width=0.45\textwidth]{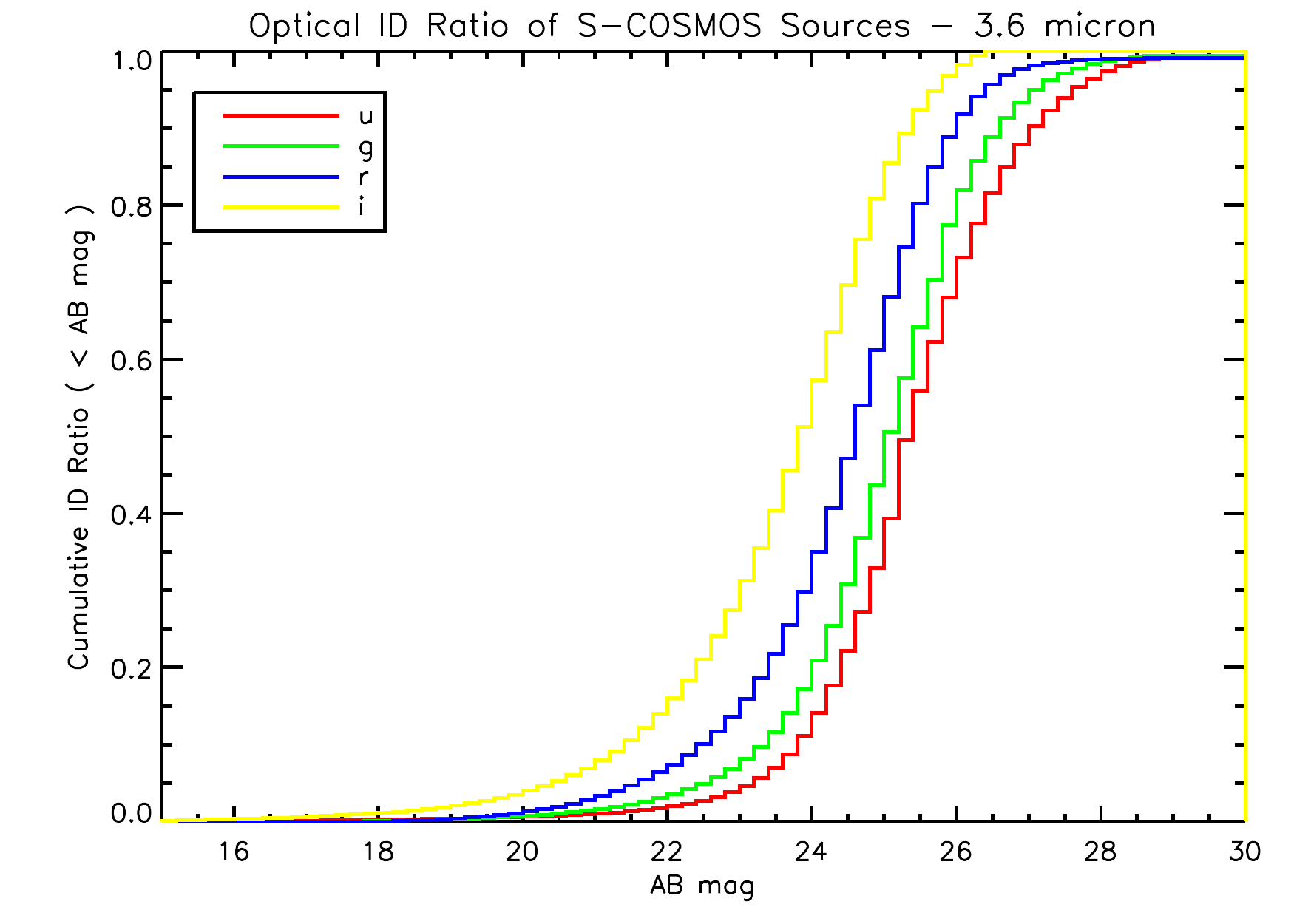}
\includegraphics[width=0.45\textwidth]{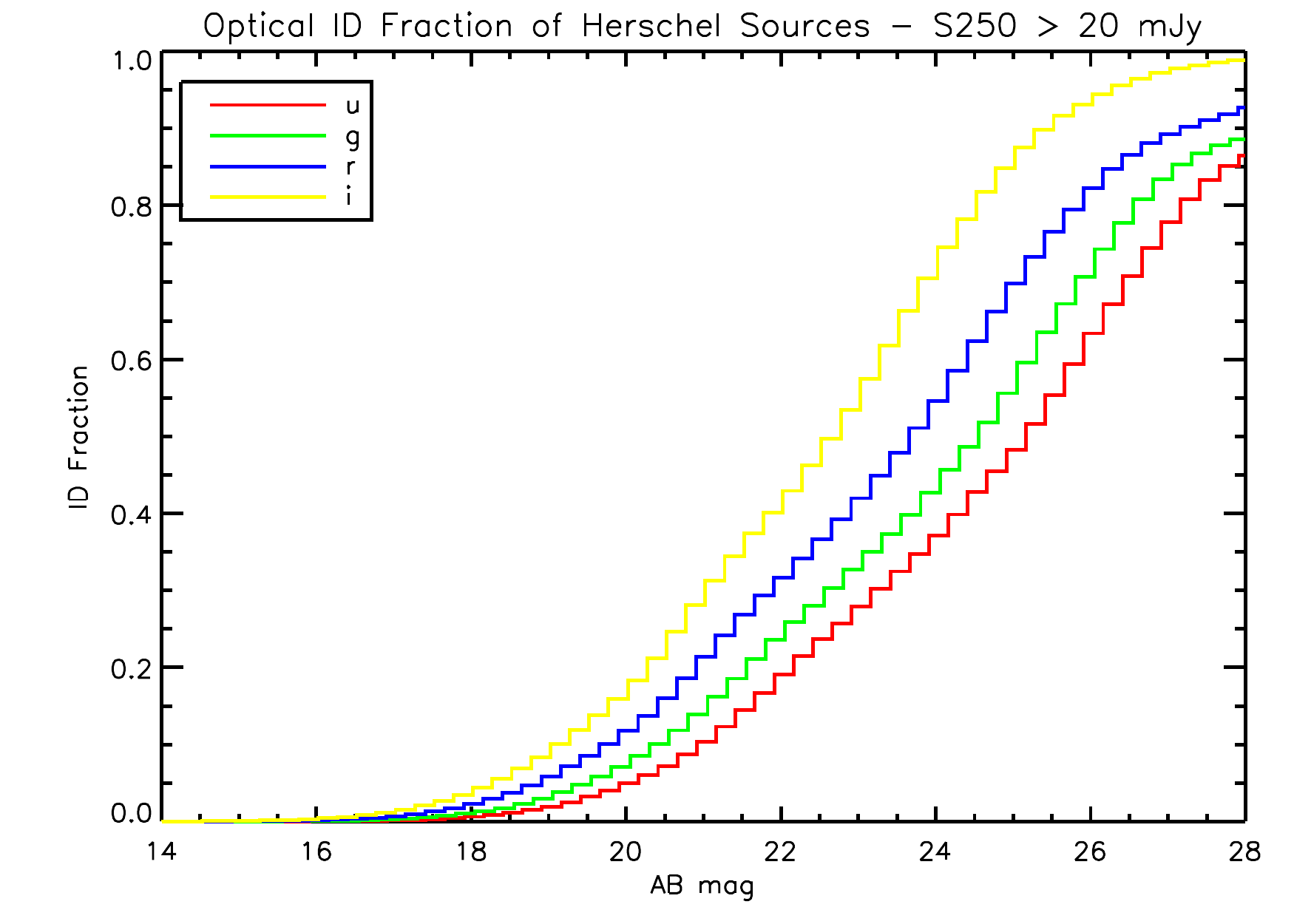}
\caption{Spitzer and Herschel IDs with VOICE Observations. \textbf{Left} : Envisaged Optical Identification Rate of Spitzer 3.6 $\mu$m sources brighter than $S_{3.6} > 2~\mu$Jy based on the COSMOS2015 Multi-Wavelength Catalogue by \cite{Laigle2016} \textbf{Right} : Envisaged Optical Identification Rate of Herschel 250 $\mu$m sources brighter than $S_{250} > 20~$mJy based on models by \cite{Xu2001}.}
\label{ids.fig}
\end{figure*}

\begin{figure*}
\centering
\includegraphics[width=0.45\textwidth]{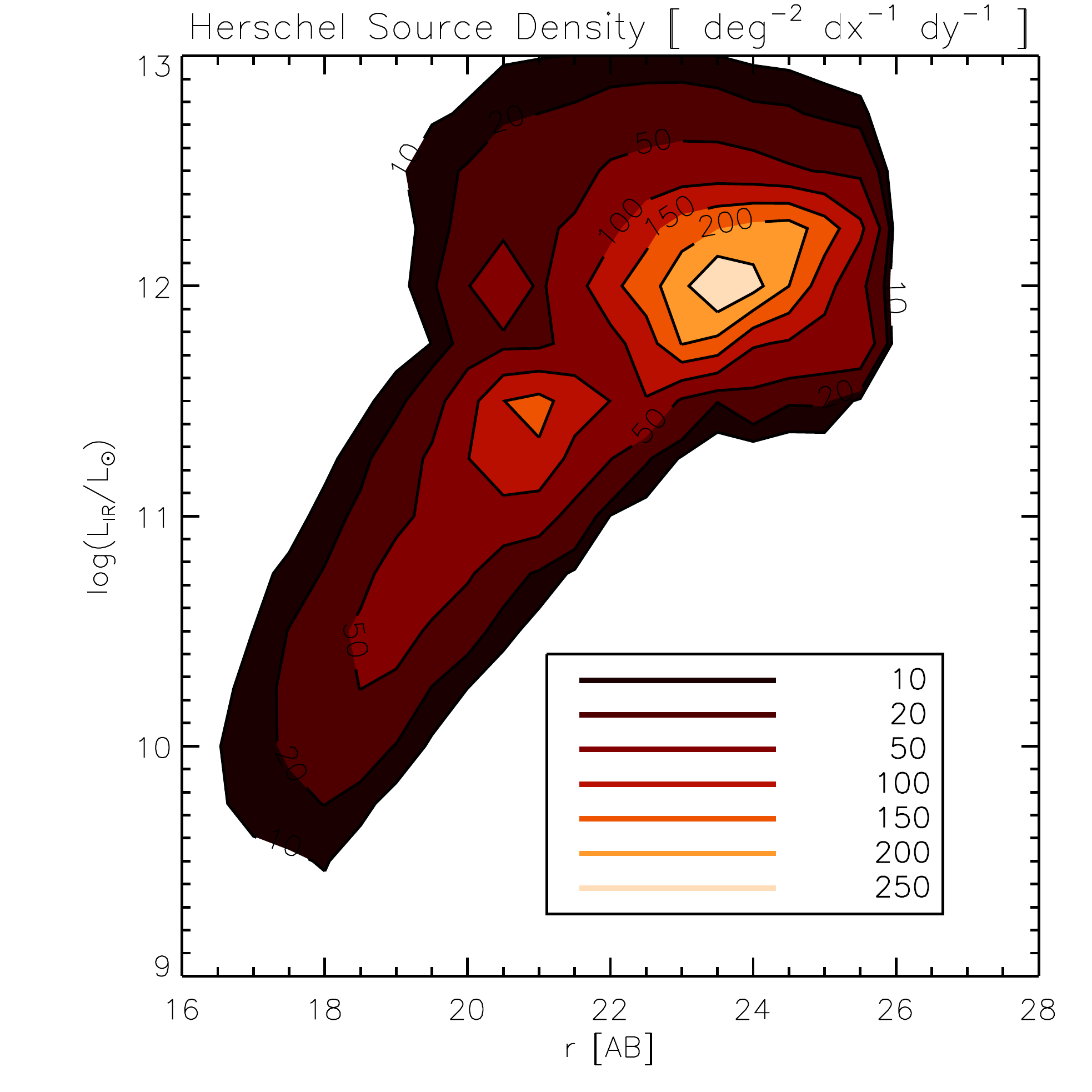}
\includegraphics[width=0.45\textwidth]{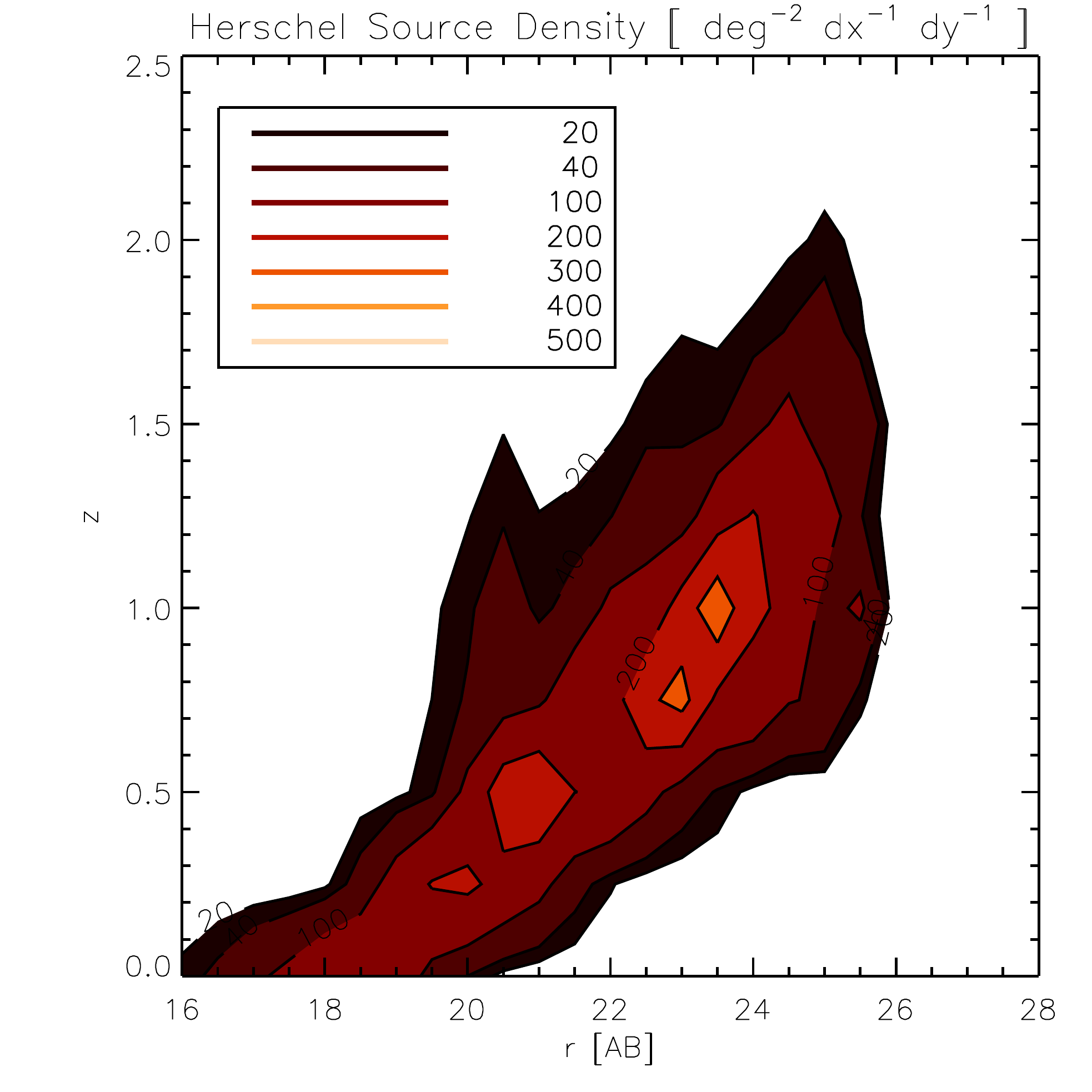}
\caption{Expected surface density of Herschel 250 $\mu$m sources brighter than $S_{250} > 20~$mJy and $r_{AB} < 26$ as a function of $r$-band magnitude (Left) and redshift (Right) based on the models by \cite{Xu2001}.}
\label{herschel.fig}
\end{figure*}

As the VST scientific operations were getting underway, we thus decided to undertake a full re-reduction of the ESIS WFI $BVR$ data, including the data already presented by \cite{Berta2006}. While this effort was originally motivated by the desire to best exploit archival optical data in the field, it has also allowed us to refine the VST-Tube \cite{Grado2012} data reduction pipeline being developed for VST surveys. Upon reduction of the full $BVR$ dataset, we decided to focus our VST observations toward obtaining deep $ugri$ optical imaging of the CDFS field while obtaining additional $u$-band imaging of the ES1 field. In order to maximize the efficient use of the INAF Guaranteed Time and strengthen the scientific synergy of the two projects, we also joined forces with the SUDARE project \cite{Cappellaro2015}, whose main aim was to carry out a supernova search \cite{Botticella2017} and study AGN variability \cite{DeCicco2015,Falocco2015} with VST. As part of this collaboration, SUDARE obtained multi-epoch $gri$ data within VOICE-CDFS, while VOICE obtained $u$-band observations and additional $gri$ data over the same field.
The VOICE-CDFS and VOICE-ES1 sky regions are covered with four VST pointings each, with a total 8 deg$^2$ sky area, as illustrated in Fig.~\ref{fields.fig}. The depth of the survey has been designed to obtain a secure optical identification of the vast majority of the SERVS and HerMES sources in at least a couple of redder optical bands as well as upper limits in the bluer optical bands. In Figure~\ref{ids.fig}, we show the envisaged fraction of Spitzer and Herschel optically-identified sources as a function of optical depth in each VST band. The planned VOICE depth at $m_{AB}\sim26$  ($5\sigma$ in a $2"$ aperture) allows to identify optical counterparts of $\> 80$\% of SERVS (S$_{3.6} > 2\mu$Jy) and HerMES (S$_{250} > 20$mJy) sources in the $r$ and $i$ bands. Figure~\ref{herschel.fig} goes on to show the envisaged surface density of Herschel sources we will be able to identify with VOICE. The approximate observing times obtained with WFI and with VST over the different pointings are summarized in Tables~\ref{esis_exp.tab} and~\ref{voice_exp.tab} respectively. 
Finally, a small amount of VOICE observations were additionally carried out within 1 deg$^2$ of the Akari Deep Field South (ADFS, also known as Akari South Ecliptic Pole field, or Akari-SEP) to support the exploitation of the Spitzer-IRAC/MIPS Extragalactic Survey (SIMES, \cite{Baronchelli2016}) as well as of deep Akari photometric surveys in the field. 

\begin{table}
\caption{ESIS WFI \protect{$30^\prime \times 30^\prime$} pointings within the ES1 field and number of 300~s exposures obtained in different filters with seeing better than 1.6 arcsec.}
\label{esis_exp.tab}
\centering
\begin{tabular}{c c c c c c}
\hline\hline
field & B & V & R & RA & Dec \\
\hline
  01 & 29 & 30 & 60 & 00:38:30.0 & -43:43:00 \\
  02 & 25 & 30 & 31 & 00:38:26.4 & -43:15:00 \\
  03 & 27 & 25 & 37 & 00:35:52.0 & -43:15:00 \\
  04 & 35 & 20 & 28 & 00:35:54.4 & -43:43:00 \\
  05 & 34 & 26 & 31 & 00:33:17.6 & -43:15:00 \\
  06 & 26 & 27 & 25 & 00:33:18.8 & -43:43:00 \\
  07 & 26 & 24 & 32 & 00:33:20.0 & -44:11:00 \\
  08 & 13 & 08 & 09 & 00:35:56.9 & -44:11:00 \\
  09 & 19 & 05 & 15 & 00:38:33.8 & -44:11:00 \\
  10 & 33 & 18 & 21 & 00:41:10.6 & -44:11:00 \\
  11 & 29 & 24 & 31 & 00:41:05.6 & -43:43:00 \\
  12 & 32 & 23 & 25 & 00:41:00.8 & -43:15:00 \\
  13 & 15 & 11 & 11 & 00:40:56.1 & -42:47:00 \\
  14 & 09 & 05 & 10 & 00:38:22.9 & -42:47:00 \\
  15 & 10 & 04 & 19 & 00:35:49.7 & -42:47:00 \\
  16 & 34 & 25 & 30 & 00:33:16.4 & -42:47:00 \\
  17 & 11 & 04 & 37 & 00:33:21.3 & -44:39:00 \\
  18 & 09 & 15 & 14 & 00:35:59.4 & -44:39:00 \\
  19 & 27 & 29 & 32 & 00:38:37.6 & -44:39:00 \\
  20 & 36 & 33 & 30 & 00:41:15.7 & -44:39:00 \\
  21 & 30 & 28 & 25 & 00:43:53.8 & -44:39:00 \\
  22 & 13 & 13 & 15 & 00:43:47.4 & -44:11:00 \\
\hline 
\end{tabular}
\end{table}
\begin{table*}
\caption{VOICE VST \protect{$1^\circ \times 1^\circ$} pointings within the CDFS, ES1 and ADFS fields and envisaged number of VST 1~hr exposures with seeing better than 1.2 arcsec. Additional $gri$ observations obtained as part of the SUDARE project are not taken into account. The envisaged depth of VOICE CDFS $ugri$ SUDARE+VOICE stacks is $m_{AB}\sim26$ ($5\sigma$ in a $2"$ aperture).}
\label{voice_exp.tab}
\centering
\begin{tabular}{c c c c}
\hline\hline
Field & RA & DEC & VST $ugriz$ [hr] \\
\hline
VOICE-CDFS-1  & 03:33:34.5  & -27:34:10.8  & 52140 \\
VOICE-CDFS-2  & 03:29:02.7  & -27:34:00.7  & 52140 \\
VOICE-CDFS-3  & 03:29:01.2  & -28:34:15.0  & 52140 \\
VOICE-CDFS-4  & 03:33:35.2  & -28:34:30.8  & 52140 \\
\hline
VOICE-ES1-1   & 00:39:25.0  & -43:30:05.0  & 50000 \\
VOICE-ES1-2   & 00:34:16.3  & -43:29:48.6  & 50000 \\
VOICE-ES1-3   & 00:34:12.5  & -44:26:34.6  & 50000 \\
VOICE-ES1-4   & 00:39:25.0  & -44:26:30.0  & 50000 \\
\hline
VOICE-ADFS-1  & 04:44:35.54 & -53:17:29.9 & 02022 \\
\hline
\end{tabular}
\end{table*}
%
%
%
\section{Data Reduction}\label{reduction.sec}
VST and WFI have been reduced using VST-Tube \cite{Grado2012}, a data reduction pipeline specifically designed for wide-field mosaiced images. VST-Tube was extensively tested and applied to WFI images before being applied to several VST/OmegaCAM Guaranteed Time Projects. In the following we describe the steps taken to reduce the VST observations of the CDFS and ES1 fields as well as the WFI observations of the ES1 field, and refer to \cite{Grado2012} for a more detailed description of the data reduction process.

Bias subtraction and flat-fielding were performed using standard procedures. High spatial frequency components are derived from twilight flat-field while the low frequency component are obtained from a combination of scientific images, we call it Skyflat, acquired preferably in the same night with a comparable exposure time. In our data reduction model we require that the photometric zero point is the same over the full field of view of each image. To this aim, we apply a gain harmonization procedure, which finds the relative CCD's gain coefficient that minimize the background differences in adjacent CCDs. In this procedure, each $CCD_{i}$ of the SkyFlat is divided in sub-windows, and in each sub-window the background is estimated using the formula: 

\begin{equation}
backg_{i} = Mode_i = 3 \times Median_i \times Mean_i
\end{equation}

The minimum $min(backg_\{i\})$ of those evaluated is assumed as background value for the $CCD_i$. The gain harmonization factor, for which the MasterFlat is then multiplied, is estimated as:

\begin{equation}
 gain_{harm-fact_i} = \frac{min(backg_i)}{median(min(backg_i))}
\end{equation}

Another correction we have to consider is the scattered light in telescope and instrument due to insufficient baffling that may cause an uncontrolled redistribution of light. In the presence of this additive effect the flat field will erroneously  describe the spatial detector response. In other words, after flat-fielding, the image background will appear flat but the photometric response will be position dependent \cite{Andersen1995}. This photometric error due to a mixing of additive an multiplicative factors can be mitigate through the determination and application of the illumination correction map (IC). The IC map, that is multiplied to the master flat field, is obtained comparing, in equatorial fields, the magnitude of observed stars with the corresponding Stetson PSF magnitudes. The resulting delta magnitude $\Delta m(x, y)$ as a function of the position is fitted with a fifth order Chebyshev polynomial surface. In order to have a flat background the IC surface, properly rescaled, is also subtracted from each images.

In VST-Tube, the absolute photometric calibration is computed on equatorial photometric standard stars fields observed each night. We compare the observed magnitude of standard stars with Stetson magnitudes \footnote{See http://www.cadc-ccda.hia-iha.nrc-cnrc.gc.ca/en/community/STETSON/standards/}. For each night and band zero-point, extinction coefficient and color term are obtained using the Photcal tool \cite{Radovich2004}. Relative photometric correction among exposures is obtained minimizing the quadratic sum of differences in magnitude between overlapping detections. The tool used for such task is SCAMP \cite{SCAMP}. Absolute and relative astrometric calibration is performed using SCAMP as well. The minimization for the relative photometry and for the relative and absolute astrometric calibration is performed in one pass on all the images covering the field. Finally, the resampling for the application of the astrometric solution and final images coaddition is applied using SWARP \cite{SWARP}.

Together with the final coadded image, for each band also we also produced a weight map that includes cold and hot pixels, cosmic rays that are detected on each exposure using SExtractor \cite{Bertin1996,SEXTRACTOR} with a specific neural network retina to identify sub-PSF sources, all multiplicative factors due to flat-field, illumination correction, gain harmonization, zero-point  and finally contains information on the number of pixels that contribute to the final mosaic.
%

Source extraction is carried out on the co-added mosaics using SEP \cite{SEP}, a python implementation of some of the algorithms in SExtractor. Sources are extracted on the $r$-band images and then measured at the positions of the $r$-band detections in other bands. Stars are identified on the basis of their PSF and the pipeline photometric calibration is improved by comparing observed stellar colors with stellar models by \cite{Girardi2005}.
\section{Multi-Wavelength Datasets}\label{multilambda.sec}
Most of the multi-wavelength data available over the VOICE CDFS and ES1 fields has been made publicly available somewhat recently and was thus not available as an homogeneous database for our study. We thus collected, re-analyzed where necessary and merged most existing multi-wavelength data ourselves so as to fully exploit it for our purposes. Available data over the vast majority of the VOICE-CDFS and VOICE-ES1 4~deg$^2$ area include the following:
\begin{enumerate}
\item GALEX UV Deep Imaging Survey \cite{Martin2005}
\item VOICE CDFS ($ugri$) and ES1 ($uBVR$) Imaging (this work)
\item VIDEO VISTA $ZYJHK$ Imaging \cite{Jarvis2013}, 
\item SERVS Spitzer-Warm 3.6 and 4.5 $\mu$m Deep Imaging \cite{Mauduit2012}
\item SWIRE Spitzer-Cold IRAC and MIPS 7-band (3.6, 4.5, 5.8, 8.0, 24, 70, 160 $\mu$m) Imaging \cite{Lonsdale2003}
\item HerMES Herschel PACS and SPIRE Imaging (100, 160, 250, 350, 500 $\mu$m) \cite{Oliver2012,Smith2012,Wang2014}
\item ATLAS 1.4~GHz Radio Continuum \cite{Franzen2015}
\end{enumerate}
While public data products are available for most of the above multi-wavelength surveys, SERVS and SWIRE data were re-extracted and band-merged with all other datasets as part of the Spitzer Data Fusion project (\cite{Vaccari2010}, \cite{Vaccari2015}, \cite{Marchetti2016}, \url{http://www.mattiavaccari.net/df/}).

The central portion of the VOICE-CDFS field, i.e. the original CDFS, was targeted by a series of increasingly deep Chandra and XMM-Netwon surveys, currently reaching a total exposure time of 7 Ms with Chandra over 0.13 deg$^2$ \cite{Luo2017} and 3.45 Ms with XMM-Newton over 0.25 deg$^2$ \cite{Comastri2011}. Similarly, only the central portion of the VOICE-ES1 field has been targeted by shallower X-ray surveys covering 0.6 deg$^2$ with both Chandra and XMM-Newton \cite{Puccetti2006, Feruglio2008}. X-ray data is thus currently available only over a small portion of the VOICE footprint. However, the ongoing X-SERVS project \cite{Brandt2016} eventually aims to produce a 12 deg$^2$ Sensitive Public X-ray Survey overing all of VOICE-CDFS and VOICE-ES1 as well as the SERVS XMM-LSS field.
%
%

\begin{figure*}
\centering
\includegraphics[width=\textwidth]{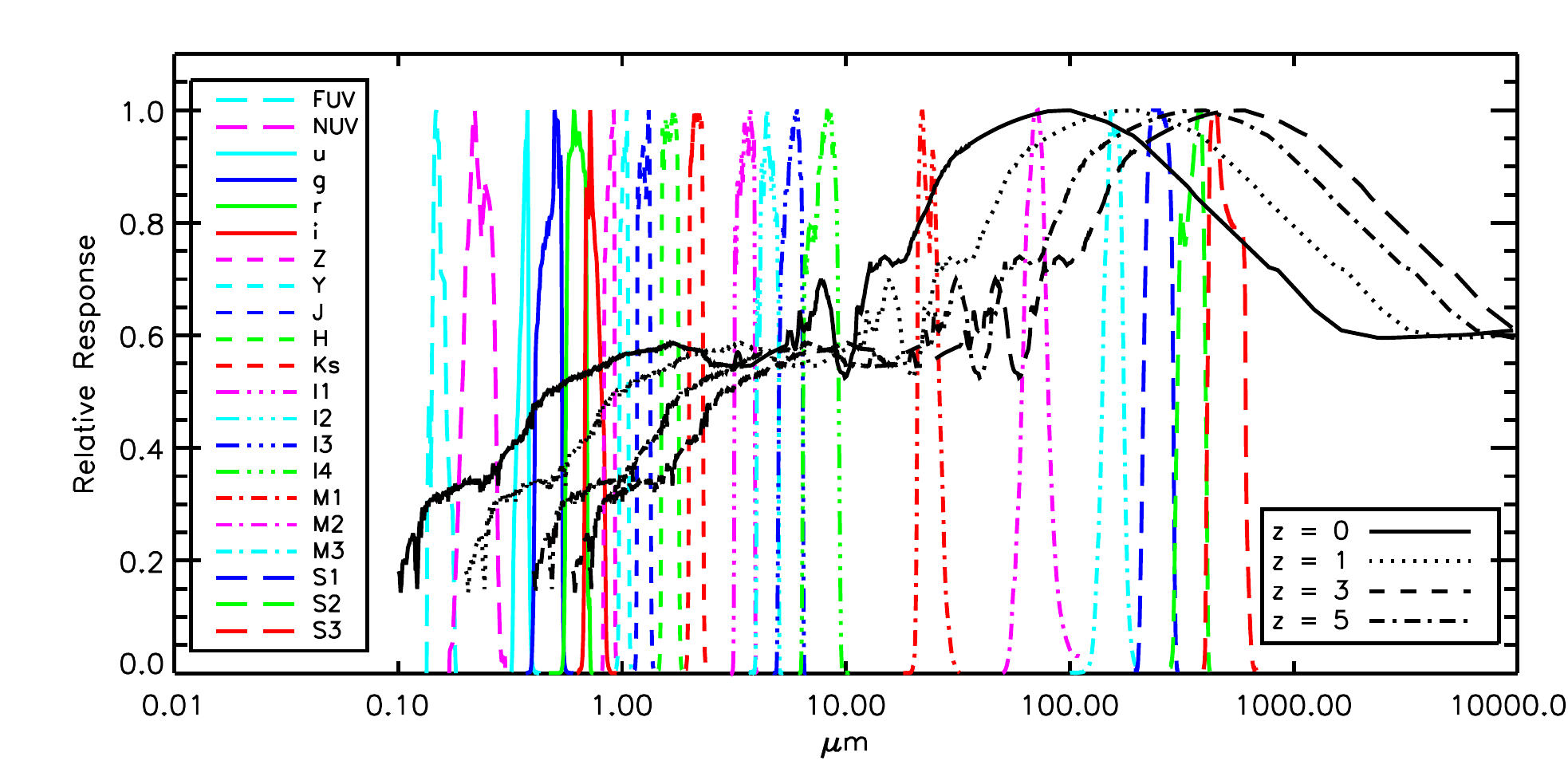}
\caption{Multi-Wavelength Coverage of Galaxy SEDs over the VOICE fields. The Arp220 SED template by \cite{Polletta2007} at different redshifts overlaid on bandpasses provided by GALEX, VST, VISTA, IRAC, MIPS and SPIRE. The redshift and dust emission of galaxies can be effectively constrained up to high redshifts.}
\label{filters.fig}
\end{figure*}

The very wide spectral coverage provided by available multi-wavelength deep imaging within the VOICE fields will allow us to characterize the properties of detected galaxies in great detail. This will build upon ongoing efforts within the SERVS collaboration to obtain photometric redshifts (Pforr et al., in prep) and obtain aperture-matched multi-wavelength photometry (Nyland et al., in prep). Firstly, this will allow us to determine accurate photometric redshifts and associated errors combining SED fitting and Machine Learning techniques as e.g. done by \cite{Cavuoti2017a,Cavuoti2017b}. Secondly, it will allow us to determine accurate physical properties such as dust attenuation, stellar masses and star formation rates using spectral modeling tools such as CIGALE \cite{Noll2009,CIGALE}. Figure~\ref{filters.fig} illustrates how the redshift as well as the peak of dust emission of a typical starburst galaxy can thus be effectively constrained up to high redshifts.
%
%
%
\section{Conclusions}\label{conclusions.sec}
We presented the VOICE survey, an optical imaging survey of the CDFS and ES1 fields. When combined with VISTA, Spitzer, Herschel and radio continuum observations over the same fields, the VOICE survey provide an invaluable resource for multi-wavelength galaxy formation and evolution studies and pave the way to multi-wavelength deep surveys with the LSST, MeerKAT and the SKA. The catalogues and their future updates will be released at \url{http://www.mattiavaccari.net/voice/} and on CDS/VizieR. They will be further expanded and refined as part of the Herschel Extragalactic Legacy Project (HELP, \cite{Vaccari2016}, \url{http://herschel.sussex.ac.uk}), an EC-REA FP7-SPACE project bringing together multi-wavelength surveys in Herschel extragalactic fields.
\section*{Acknowledgements}
Mattia Vaccari is supported by the European Commission's REA (FP7-SPACE-2013-1 GA 607254 - Herschel Extragalactic Legacy Project), South Africa's DST (DST/CON 0134/2014) and Italy's MAECI (PGR GA ZA14GR02 - Mapping the Universe on the Pathway to SKA).
%


\newcommand{\aj}{AJ}
\newcommand{\araa}{ARA\&A}
\newcommand{\apj}{ApJ}
\newcommand{\apjl}{ApJ} 
\newcommand{\apjs}{ApJS} 
\newcommand{\ao}{Appl.~Opt.}
\newcommand{\apss}{Ap\&SS}
\newcommand{\aap}{A\&A}
\newcommand{\aapr}{A\&A~Rev.}
\newcommand{\aaps}{A\&AS}
\newcommand\baas{\ref@jnl{BAAS}}%
\newcommand{\memsai}{Mem.~Soc.~Astron.~Italiana}
\newcommand{\mnras}{MNRAS}
\newcommand{\nat}{Nature}
\newcommand{\pasp}{PASP}
\newcommand{\pasj}{PASJ}
\newcommand{\skytel}{S\&T}
\newcommand{\procspie}{Proc.~SPIE}
\bibliographystyle{JHEP}
\bibliography{voice}
\end{document}